\documentclass[aps,prl,twocolumn,superscriptaddress,sort&compress]{revtex4-1}%
\usepackage[T1]{fontenc}
\usepackage{eurosym}
\usepackage{amsfonts}
\usepackage{amssymb}
\usepackage{amsmath}
\usepackage{floatrow}
\usepackage[position=t,singlelinecheck=off,caption=false]{subfig}
\usepackage{graphicx}
\usepackage{xcolor}%
\usepackage{booktabs}
\usepackage{adjustbox}
\usepackage[hyperfigures=true,bookmarksnumbered=true,bookmarksopen=true,bookmarks=true,linkcolor=blue,urlcolor=blue,citecolor=blue,colorlinks=true,pdfhighlight=/O,pdfstartview={XYZ null null 1}]{hyperref}
\graphicspath{{}{images/}}
\providecommand{\U}[1]{\protect\rule{.1in}{.1in}}
\providecommand{\U}[1]{\protect\rule{.1in}{.1in}}
\def\HG#1 {\emph{\color{blue}#1}}
\def\CFA {CaFe$_{2}$As$_{2}$}
\def\CFRALOW {Ca--Rh$0.28$}
\def\CSFRALOWHIGH {CaSr--Rh0.25\ and\ CaSr--Rh0.48}

\def\CSFRAHIGH {CaSr--Rh$0.48$}

\begin{document}
\title{The collapsed tetragonal phase as a strongly covalent and fully nonmagnetic state: persistent magnetism with interlayer As--As bond formation in Rh-doped Ca$_{0.8}$Sr$_{0.2}$Fe$_2$As$_2$}
\author{K. Zhao}
\affiliation{Experimentalphysik VI, Center for Electronic Correlations and Magnetism, University of Augsburg, 86159 Augsburg, Germany}
\author{J. K. Glasbrenner}
\affiliation{Department of Computational and Data Sciences/Computational Materials Science Center, George Mason University, 4400 University Drive, Fairfax, VA 22030}
\affiliation{National Research Council/Code 6393, Naval Research Laboratory, Washington DC 20375, USA}
\author{H. Gretarsson}
\affiliation{Max-Planck-Institut f\"{u}r Festk\"{o}rperforschung,
Heisenbergstra$\mathrm{\beta}$e 1, D-70569 Stuttgart, Germany}
\author{D. Schmitz}
\affiliation{CPM, Institute of Physics, University of Augsburg, 86135 Augsburg, Germany}
\author{J. Bednarcik}
\author{M. Etter}
\affiliation{Deutsches Elektronen-Synchrotron (DESY) 22607 Hamburg Germany}
\author{J. P. Sun}
\affiliation{Beijing National Laboratory for Condensed Matter Physics, and Institute of Physics, Chinese Academy of Sciences, Beijing 100190, China}
\author{R. S. Manna}
\affiliation{Experimentalphysik VI, Center for Electronic Correlations and Magnetism, University of Augsburg, 86159 Augsburg, Germany}
\author{A. Al-Zein }
\author{S. Lafuerza}
\affiliation{European Synchrotron Radiation Facility, BP 220, F-38043 Grenoble Cedex, France}
\author{W. Scherer}
\affiliation{CPM, Institute of Physics, University of Augsburg, 86135 Augsburg, Germany}
\author{J. G. Cheng}
\affiliation{Beijing National Laboratory for Condensed Matter Physics, and Institute of Physics, Chinese Academy of Sciences, Beijing 100190, China}
\author{P. Gegenwart}
\affiliation{Experimentalphysik VI, Center for Electronic Correlations and Magnetism, University of Augsburg, 86159 Augsburg, Germany}

\begin{abstract}
A well-known feature of {\CFA}-based superconductors is the pressure-induced collapsed tetragonal phase that is commonly ascribed to the formation of an interlayer As--As bond. 
Using detailed X-ray scattering and spectroscopy, we find that Rh-doped Ca$_{0.8}$Sr$_{0.2}$Fe$_{2}$As$_{2}$ does not undergo a first-order phase transition and that local Fe moments persist despite the formation of interlayer As--As bonds. Our density functional theory calculations reveal that the Fe--As bond geometry is critical for stabilizing magnetism and that the pressure-induced drop in the $c$ lattice parameter observed in pure {\CFA} is mostly due to a constriction within the FeAs planes. These phenomena are best understood using an often overlooked explanation for the equilibrium Fe--As bond geometry, which is set by a competition between covalent bonding and exchange splitting between strongly hybridized Fe $3d$ and As $4p$ states. In this framework, the collapsed tetragonal phase emerges when covalent bonding completely wins out over exchange splitting. 
Thus the collapsed tetragonal phase is properly understood as a strong, covalent phase that is fully nonmagnetic with the As--As bond forming as a byproduct.
\end{abstract}
\maketitle

The pressure-induced collapsed tetragonal (CT) phase transition \cite{Ames_2008, Ames_2009, canfield_2009} of the iron-based superconductor {\CFA} \cite{NINI_2008, Ronning_2008}  is a structural transition characterized by a discontinuous change in the material's lattice parameters and volume. The transition is unique among the ThCr$_{2}$Si$_{2}$ (122) structural family of iron-based superconductors~\cite{Hosono_2008, XHChen_2008, NLWang_2008, Rotter_2008, Jeevan_2008, CQJin_2008, FeSe_2008, Review_2010, Mazin_2011}, occurring at a hydrostatic pressure of 0.35 GPa \cite{Ames_2008} that is an order of magnitude lower than the continuous (second order) transitions observed in the other members of the AFe$_{2}$As$_{2}$ (A = Ba, Sr, Eu) family~\cite{Ba122CT_2010, Ba122CT_2011}. The CT phase itself is nonmagnetic, lacks magnetic fluctuations \cite{Ames_2009, CTNeutron_2013}, exhibits Fermi liquid behavior \cite{CaP122_2011, CaRh122_2011}, and is not superconducting \cite{yu_2009}, which supports a spin-fluctuation model of superconductivity. There is a diversity of opinion on how to describe the Fe moment for the CT phase transition, with most models belonging to one of three categories: (1) the magnetism is itinerant and the Fe moment is quenched when a Fermi surface nesting vector disappears due to pressure \cite{zhang_2009}, (2) the magnetism is local and the Fe moment is quenched when pressure-induced gains in the Gibb's free energy win out over the Hund's coupling \cite{ji_2011}, and (3) each Fe$^{2+}$ site has six $3d$ electrons arranged in one of three distinct spin states, $S=0$ (nonmagnetic), $S=1$ (low spin), and $S=2$ (high spin), and applying pressure transitions a majority of the Fe sites from $S=2$ to $S=0$ or $S=1$, suppressing magnetism \cite{Hlynur_2013}. Regardless of the way one models the Fe magnetic moment, the driving mechanism of the CT phase is generally attributed to a well-known feature of the CT phase, the strong interlayer As--As covalent bond \cite{CTDFT_2009}. Stronger interlayer As--As bonds will promote smaller interlayer As--As bond lengths as Hoffman and Zheng showed in their bond analysis of the ThCr$_{2}$Si$_{2}$ structural compounds \cite{hoffman_and_zheng}, and thus the CT phase transition occurs when the As--As bond length decreases below a critical value of 3 {\AA} \cite{CaSr122_2012}, at which point the As--As bonding energy wins out over the magnetic energy and induces a first-order structural transition that quenches the Fe moments.

\floatsetup[figure]{style=plain,subcapbesideposition=top}
\begin{figure*}[th!]
  \sidesubfloat[]{\includegraphics[width=0.13\textwidth]{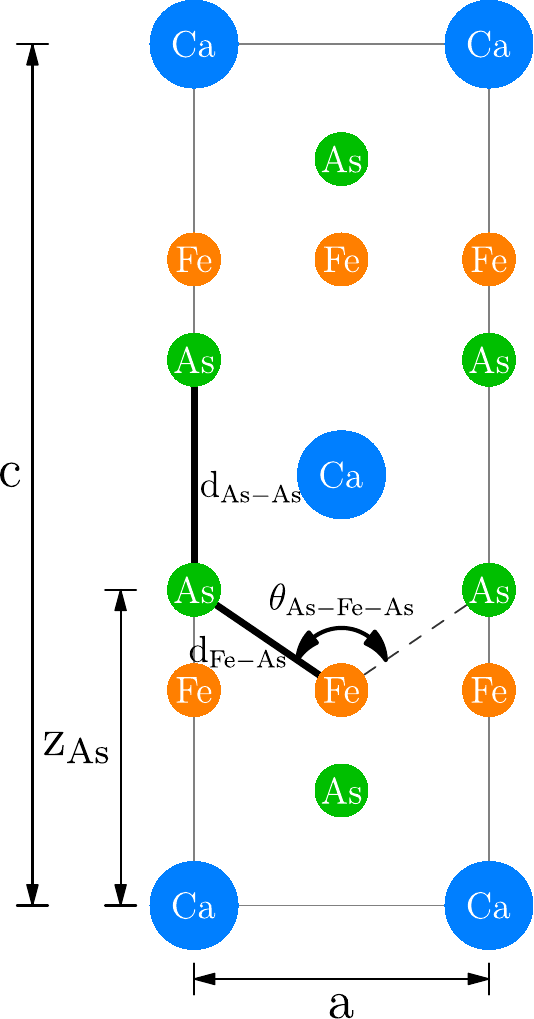}\label{fig:cfa_structure}}\quad
  \sidesubfloat[]{\includegraphics[width=0.39\textwidth]{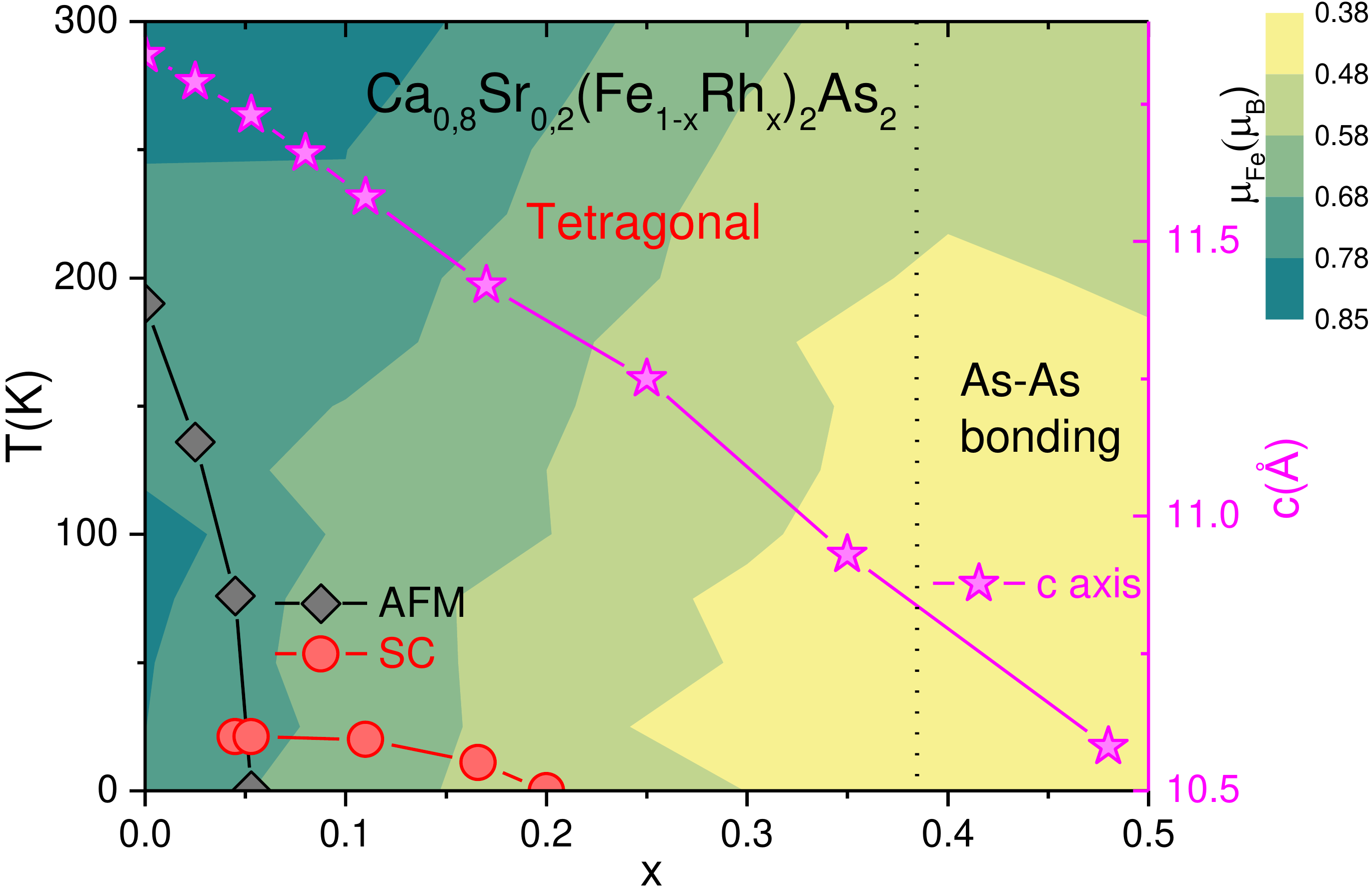}\label{Phase_diagram}}\quad
  \sidesubfloat[]{\includegraphics[width=0.32\textwidth]{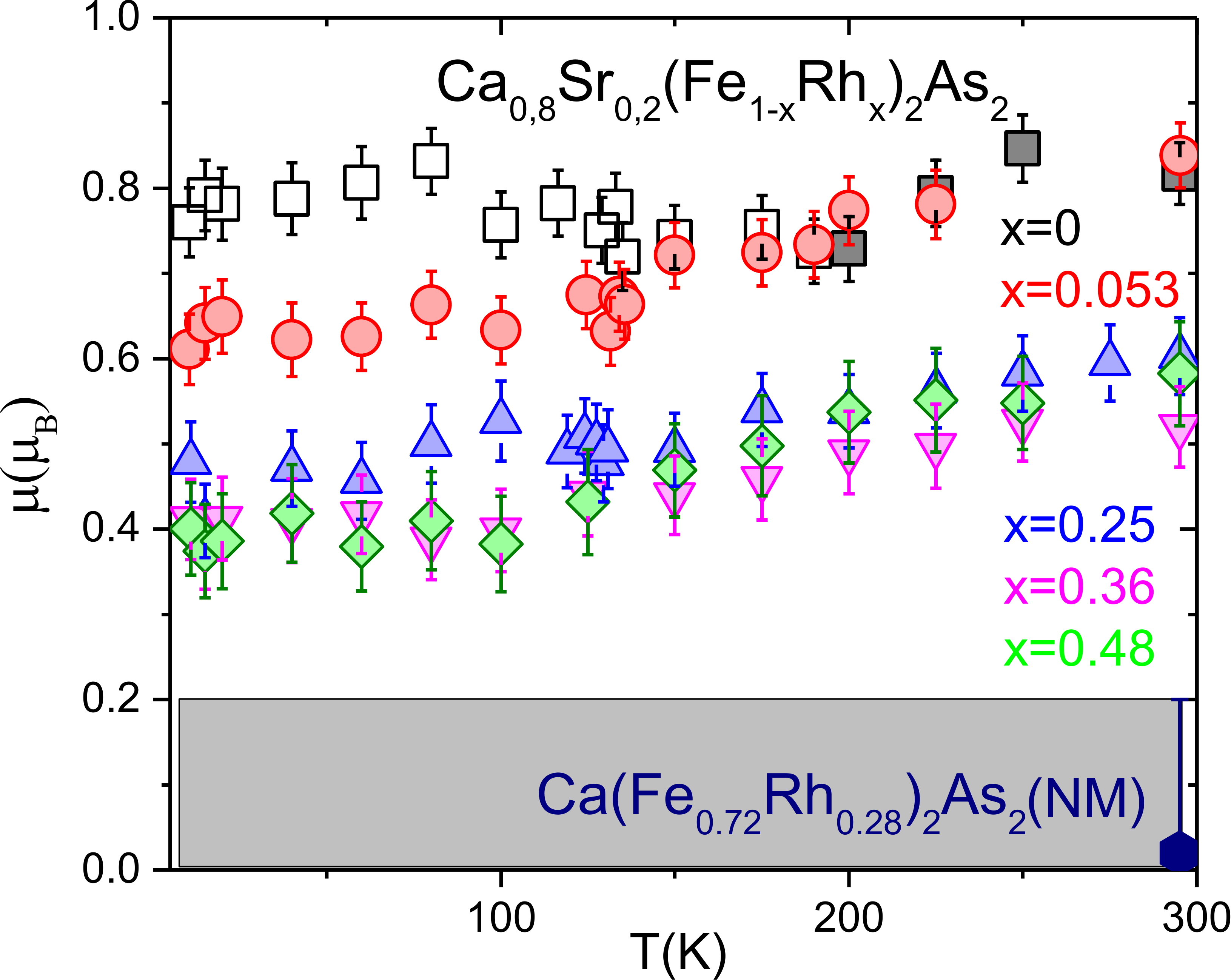}\label{XES_and_PDF}}
  \caption{(a) Sideview of the crystal structure of {\CFA} with
    labels for the structural parameters. (b) Temperature composition $x$ phase
    diagram for Ca$_{0.8}$Sr$_{0.2}$(Fe$_{1-x}$Rh$_x$)$_2$As$_2$ with size of
    Fe moments as determined from x-ray emission spectroscopy. The diamonds and
    filled circles indicate the antiferromagnetic and superconducting
    transition, respectively, determined from electrical resistivity and
    thermal expansion. The pink stars indicate the size of the c-lattice
    parameter at room temperature (right y-axis) which decreases linearly to
    values characteristic for the interlayer As--As bonding beyond $x=0.38$,
    cf. cyan dotted vertical line. (c) Temperature dependence of the Fe local
    moment $\mu$
    ($\mu_{B}$) for
    Ca$_{0.8}$Sr$_{0.2}$(Fe$_{1-x}$Rh$_x$)$_2$As$_2$ (x=0, 0.053, 0.25, 0.36,
    and 0.48) derived from respective XES spectra, as described in text. The
    open square represents local moment in antiferromagnetic order state. The
    local moment for Ca(Fe$_{1-x}$Rh$_x$)$_2$As$_2$
    ($x=0.28$) is also shown for comparison.}
  \label{fig:structure_and_measurement}
\end{figure*}

These models face challenges when applied to {\CFA}-based chemical substitution experiments \cite{jeffries_2014, knoner_2016, CaRh122_2011, Kan_2015, Saha_2012, CaPrPNAS_2011, CaPrSTM_2014}. For example, substituting 33\% of Ca sites with Sr and applying pressure leads to a paramagnetic CT phase (defined here as a structure with an As--As bond length shorter than 3 {\AA}) instead of a nonmagnetic one, with a transition that may be second order \cite{jeffries_2014}. A pure itinerant model cannot explain this paramagnetic CT phase, while the challenge for localized models is to provide an explanation for why identical cell volumes result in a paramagnetic phase for the Sr-doped case and a nonmagnetic phase for the undoped case. A localized model may be able to quantify the trend by fitting to electronic structure calculations, but this limits its explanatory power as this effectively includes itinerant features, while a mixed valence model, as acknowledged in Ref.~\onlinecite{Hlynur_2013}, cannot explain the first-order collapse to a nonmagnetic state and thus is unable to identify a mechanism for what Sr doping changes in the material. In addition, the CT phase being proximate to a high-temperature superconducting phase above 45 K in rare-earth doped {\CFA} \cite{Saha_2012, CaPrPNAS_2011, CaPrSTM_2014} illustrates the need for a comprehensive understanding of the CT phase. 

There are also clues from density functional theory (DFT) calculations that the role of the As--As bond in the CT phase transition needs to be reassessed.
A remarkable success of DFT is that it can distinguish between the uncollapsed and collapsed phases in the 122 family, as structural relaxation calculations using a magnetic structure with $\mathbf{q} = (\pi, 0)$ or $(0, \pi)$ (known as the single stripe pattern) reproduces the lattice parameters of the uncollapsed tetragonal phase, while relaxing in the nonmagnetic state reproduces the CT phase \cite{CTDFT_2009, tomic_2012, wang_2009, colonna_2011, widom_2013}. In terms of chemical bonding, both DFT \cite{widom_2013} and DFT-based dynamical mean-field theory (DFT+DMFT) \cite{diehl_2014, van_Roekeghem_2016} calculations find that there is a substantial amount of hybridization between the Fe $3d$ and As $4p$ states, which becomes \textit{stronger} in the CT phase despite the formation of the As--As bond. The substantial Fe--As hybridization was previously discussed as a general feature of the pnictides \cite{Belashchenko_2008}, but this finding usually is not factored into models of the CT phase. This would seem to be a mistake. In the magnetic, uncollapsed phase, the Fe--As bonding and antibonding hybrid bands themselves are exchange split, leading to a competition between covalent bonding and the magnetic energy reflected in the equilibrium distance between neighboring Fe and As planes. Increasing the exchange splitting weakens the covalent bond as electrons start to occupy the majority antibonding band, while reducing the exchange splitting empties the majority antibonding band and strengthens the covalent bond \cite{Belashchenko_2008}. Indeed, the proximity of such antibonding states to the Fermi level was reported in Hoffman and Zheng's analysis (here the bond is Mn--P) \cite{hoffman_and_zheng}. Furthermore, other key quantities in DFT+DMFT calculations, such as the $d_{xy}$ orbital's imaginary part of the self-energy, are quite sensitive to the Fe--As bond geometry (which changes across the CT phase transition) but not to the As--As bond length \cite{van_Roekeghem_2016}. This mounting evidence seems to suggest that the CT phase transition has less to do with As--As bond and more to do with the Fe--As bond geometry. As we'll show below, our measurements and calculations are in agreement with this hypothesis.

In this Letter, we report on the changes in the structural and magnetic properties of Sr- and Rh-doped {\CFA} using electrical resistivity, thermal expansion, X-ray diffraction (XRD) and emission spectroscopy (XES) measurements. Substituting Fe with Rh \cite{CaRh122_2011} provides chemical pressure, avoiding the challenges inherent in performing spectroscopic measurements under high pressures. We obtain the surprising result that Rh-doped Ca$_{0.8}$Sr$_{0.2}$Fe$_2$As$_2$ does not undergo a first-order phase transition as the As--As distance crosses the critical 3 {\AA} threshold \cite{CaSr122_2012} and that the local Fe moments persist despite the formation of interlayer As--As bonds. We further analyze these measurements using DFT calculations, finding that subtle variations in the Fe--As bond geometry in Sr-doped and Sr-free samples, regardless of the As--As bond length, determines whether Fe is magnetic or not, and that a first order CT phase transition corresponds to the intralayer constriction of neighboring As--Fe--As planes due to quenching of magnetism. These results show a complexity that cannot be explained in models that require a sharp distinction between low-spin and high-spin states, or that start from a fully localized or itinerant description. Instead, the same set of electrons both provide the local moments and form the Fe--As and As--As bonds, so we interpret our results using the framework of a competition between covalent bonds and exchange splitting \cite{Belashchenko_2008}. In this picture, forming a covalent As--As bond does not require the quenching of magnetism and the transition to the CT phase is allowed to be continuous, depending on the details of how the Fe--As bond geometry evolves with pressure. Furthermore, the CT phase is properly identified not by a sub-3 {\AA} As--As bond length \cite{CaSr122_2012}, but instead as a fully nonmagnetic phase with strong Fe--As and As--As covalent bonds.
This framework provides a common mechanism for the nonmagnetic CT phase in {\CFA} and the other 122 materials~\cite{Ba122CT_2010, Ba122CT_2011}.

Fig.\ \ref{Phase_diagram} displays the phase diagram for Ca$_{0.8}$Sr$_{0.2}$(Fe$_{1-x}$Rh$_x$)$_2$As$_2$ $(0 < x < 0.48)$, based on X-ray diffraction and spectroscopy, electrical resistivity and thermal expansion, see Supplemental Materials for more details \cite{SM_2016}. Upon Rh doping, bulk superconductivity emerges after the complete suppression of the antiferromagnetic orthorhombic phase with maximal transition temperature of 21 K. More Rh doping likely suppresses the antiferromagnetic fluctuations, and the superconducting phase vanishes around $x = 0.20$. 
The linear decrease of the c--axis parameter with $x$ at 300 K (cf. stars and right y--axis) indicates the absence of a first-order transition, in contrast to the sharp drop of $c$--axis for Ca(Fe$_{1-x}$Rh$_x$)$_2$As$_2$ \cite{SM_2016}, due to the CT phase transition at 300 K in the latter material~\cite{CaRh122_2011}. Based on the electrical resistivity measurements on Ca$_{0.8}$Sr$_{0.2}$Fe$_2$As$_2$ under hydrostatic pressure, a similar phase diagram compared to the Rh doped case can be constructed \cite{SM_2016}, again without a first-order phase transition.

\begin{table}
  \begin{adjustbox}{max width=0.98\textwidth}
    \begin{tabular}{ccc}
      \toprule
      300K & Ca$_{0.8}$Sr$_{0.2}$(Fe$_{1-x}$Rh$_x$)$_2$As$_2$ & Ca(Fe$_{1-x}$Rh$_x$)$_2$As$_2$ \\
           & (x=0.25, and 0.48) & (x=0.28) \\
      \midrule
           &Lattice parameter({\AA}) & \\
      \midrule
      a  & 3.9891(6)/4.06610(10) & 4.0270(3) \\
      c  & 11.2556(17)/10.6100(2) & 10.6450(9) \\
      \midrule
           & Atomic sites & \\
      \midrule
      Ca (Sr) & 2a (0, 0, 0) & 2a (0, 0, 0) \\
      Fe(Rh) & 4d (0, 0.25, 0.5) & 4d (0, 0.25, 0.5) \\
      As & 4e (0, 0, 0.36579(5)/0.36806(5)) & 4e (0, 0, 0.36763(6)) \\
      \midrule
           & Average bond lengths ({\AA}) & \\
      \midrule
      Fe--As & 2.3826(4)/2.3880(3) & 2.3711(4) \\
      Fe--Fe & 2.8207(4)/2.8752(1) & 2.8475(2) \\
      As--As & 3.0213(12)/2.7998(12) & 2.8181(13) \\
      \midrule
           & Average bond angles (deg) & \\
      \midrule
      As--Fe--As ($\theta$) & 113.68(2)/116.72(1) & 116.25(3) \\
      As--Fe--As ($\beta$) & 107.410(12)/105.971(10) & 106.194(12) \\
      \bottomrule
    \end{tabular}
  \end{adjustbox}
  \caption{Crystallographic data of
    Ca$_{0.8}$Sr$_{0.2}$(Fe$_{1-x}$Rh$_x$)$_2$As$_2$ (x=0.25, and 0.48) and
    Ca(Fe$_{1-x}$Rh$_x$)$_2$As$_2$ (x=0.28)}
  \label{table1}
\end{table}

To investigate the structural details and atomic coordinates, single crystal X-ray diffraction was conducted for Ca$_{0.8}$Sr$_{0.2}$(Fe$_{1-x}$Rh$_x$)$_2$As$_2$ ($x=0.25$ and $0.48$) [{\CSFRALOWHIGH}] and Ca(Fe$_{1-x}$Rh$_x$)$_2$As$_2$ ($x=0.28$) [{\CFRALOW}], yielding the results listed in Table~\ref{table1}. The geometry parameters of the FeAs$_4$ tetrahedra including the Fe--As bond length and As--Fe--As angle are derived together with the interlayer As--As distance. For {\CSFRAHIGH}, this As--As distance is 2.8 {\AA}, similar to the value of {\CFRALOW}. Note that both distances are less than the critical value of 3 {\AA} \cite{CaSr122_2012}. Independent pair distribution function (PDF) measurements show the same formation of short interlayer As--As bonds in {\CFRALOW} and {\CSFRAHIGH} at 300 K, see Supplemental Materials \cite{SM_2016}. Additionally the PDF measurements confirm that the Fe--As bond is enhanced for {\CSFRAHIGH} by about 0.01 {\AA} compared to {\CFRALOW}.


X-ray emission spectroscopy (XES) has emerged as a useful technique to study the fast fluctuating magnetic moments in Fe based superconductors~\cite{Hlynur_2011, Hlynur_2013, Hlynur_2015, Ba122_2016}. To investigate the link between the structure of Ca$_{0.8}$Sr$_{0.2}$(Fe$_{1-x}$Rh$_x$)$_2$As$_2$ and its fluctuating Fe moment, we measured the temperature and doping dependence of the Fe K$\beta$ emission line. By application of the integrated absolute difference (IAD) analysis on the shape of the emission line, information on the size of the Fe magnetic moment can be obtained, see Supplemental Materials for details \cite{SM_2016}. In Fig.~\ref{XES_and_PDF}, we plot the temperature dependence of the local moment for Ca$_{0.8}$Sr$_{0.2}$(Fe$_{1-x}$Rh$_x$)$_2$As$_2$ ($x=0$, 0.053, 0.25, 0.36, and 0.48), extracted from the emission line as described above. The detection limit (zero signal) of the IAD technique is shown by the shaded area~\cite{Hlynur_2013}. At room temperature the samples with lower Rh doping ($x=0$ and $x=0.053$) have a local moment around 0.8 $\mu_B$, which upon cooling gradually decreases to around 0.6 $\mu_B$. However, for $x=0$, the moment starts to increase below the N\'{e}el temperature until the room temperature value is reached again. In the $x=0.053$ sample where long-range order is suppressed no such a reversal is observed. The higher-doped samples, for which interlayer As--As bonds are formed, show a finite but reduced moment of around 0.6 $\mu_B$ at room temperature. Upon cooling, this moment also gradually decreases from $\approx 0.6$ $\mu_B$ at $T=295$ K to 0.4 $\mu_B$ at 10 K. This observation is in stark contrast to our results on {\CFRALOW} which shows a non-magnetic state at 300 K.

\begin{figure}[t]
  \includegraphics[width=\textwidth]{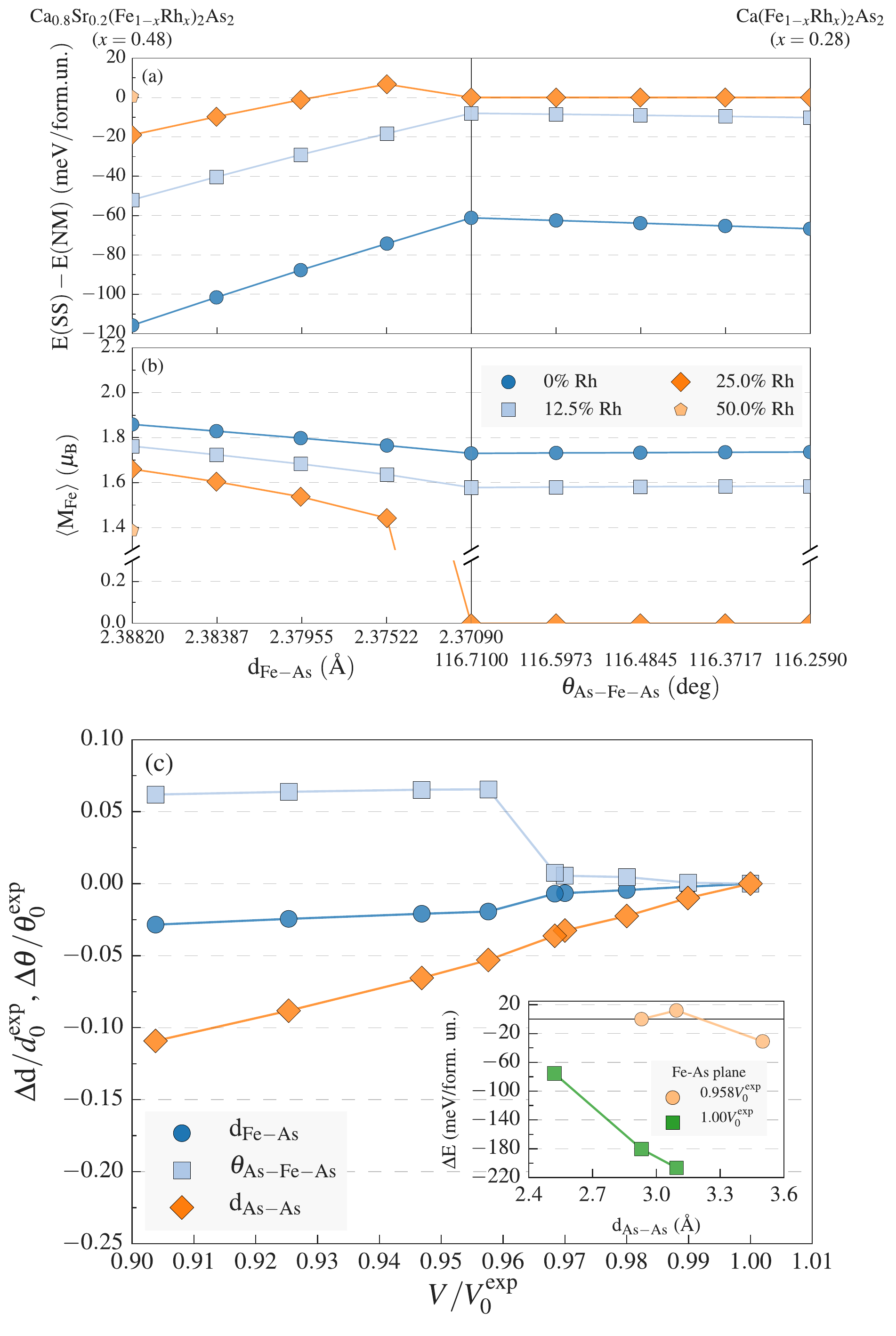}
  \caption{Density functional theory calculations of the energy difference
    between the single stripe (SS) and nonmagnetic (NM) states (panel a) and
    the average Fe local moment (panel b) as you interpolate between the
    structural parameters for {\CSFRAHIGH} (far left) and {\CFRALOW} (far
    right), starting with d$_{\text{Fe--As}}$ first and
    $\theta_{\text{As--Fe--As}}$ second. The central vertical line separates
    the two interpolation stages and the Rh doping levels are indicated in the
    legend. In panel c the optimized structural parameters of pure {\CFA} are
    plotted as a function of the volume per formula unit. The reference point
    for the fractional volume and bond lengths/angle is the optimized structure
    for the experimental volume. Inset: The energy difference
    $\Delta \text{E} = \text{E(SS)} - \text{E(NM)}$ for the system at volumes
    constrained to 95.8\% and 100\% of the experimental volume when the FeAs
    plane geometry is fixed and d$_{\text{As--As}}$ is varied.}
  \label{fig:dft_calculations}
\end{figure}


To understand why the magnetism persists for {\CSFRAHIGH} despite the As--As bond length being close in value to the pressure-induced CT phase of undoped {\CFA} (d$_{\text{As--As}} = 2.7952$ {\AA}) \cite{Ames_2008}, we performed a series of DFT calculations using the pseudopotential software package \textsc{vasp} with projector augmented wave potentials \cite{vasp1, vasp2} and the GGA exchange-correlation functional \cite{gga}, see Supplemental Materials for details \cite{SM_2016}.
We found that subtle changes in the lattice parameters due to chemical pressure affect the magnetic stability. Fixing the lattice parameters to the {\CFRALOW} values reported in Table \ref{table1} weakens the magnetic stability, and magnetism is fully suppressed with a Rh doping level of 25\% or higher. In contrast, fixing the lattice parameters to the {\CSFRAHIGH} values stabilizes magnetism, with the antiferromagnetic single stripe (SS) phase remaining stable up to 25\% Rh doping and the Fe local moment was found to be still present at 50\% Rh doping \footnote{We tested different arrangements of substitutional Rh doping at 50\%, in some configurations the local moments are suppressed, while in others the local moments persist. On average, this indicates that a lack of long-range order but does not preclude local moment formation.}.

Next, we assessed the role of the Fe--As bond in stabilizing magnetism using a second set of DFT calculations, which are shown in Fig.~\ref{fig:dft_calculations}. These calculations interpolated between the {\CSFRAHIGH} and {\CFRALOW} lattice parameters in a two-stage process (the As--As bond length was fixed to its {\CFRALOW} value). We found that in the absence of electron doping that a 0.72\% reduction in the Fe--As bond length increases the energy of the magnetic phase relative to the nonmagnetic phase by nearly a factor of 2. Furthermore, with 25\% Rh doping the system transitions to a nonmagnetic phase at a Fe--As bond length of 2.37522 {\AA} as the SS phase becomes metastable (it becomes unstable at 2.37090 {\AA}). The bond angle As--Fe--As is relatively unchanged in the interpolation, however in a separate calculation we found that increasing the angle by 1.3 degrees lowered the energy by about 10 meV. This confirms that even in this ``collapsed'' environment that small changes in the Fe--As bond geometry affect the stability of the magnetic phase.


These results show a sharp division between two phases (one that is magnetic in some way, while the other is fully nonmagnetic) that depends on the geometry of the Fe--As bond even when the As--As bond length is less than 3 {\AA}, which presents a serious challenge to the theory that forming an As--As bond \textit{drives} the CT phase transition. One possible objection to this would be that we've only shown this to hold for characteristics that might be specific to doped {\CFA}. To show that this result is more general, we performed structural relaxation calculations of undoped {\CFA} where we optimized the structure for a series of fixed volumes. The resulting bond lengths (Fe--As and As--As) and bond angle (As--Fe--As) as a function of cell volume are plotted in Fig.~\ref{fig:dft_calculations}(c), with the CT phase transition occurring between the volumes 0.958$V_{0}^{\text{exp}}$ and 0.968$V_{0}^{\text{exp}}$ ($V_{0}^{\text{exp}}$ is the experimental volume at ambient pressure \cite{Goldman_2008}), which causes a discontinuous $5.5\%$ reduction in the $c$ parameter (the $a$ parameter increases in response to compensate and preserve the fixed volume).
What hasn't been pointed out in previous discussions is that $83\%$ of the $c$ parameter's decrease is due to a change in the Fe--As bond geometry (this mostly stems from the $6\%$ increase in the bond angle), with the remaining $17\%$ is due to a decrease in the As--As interlayer distance. Put another way, the collapse of the $c$ parameter is the consequence of a \textit{sudden constriction of the interlayer distance between neighboring Fe and As planes}, which occurs when magnetism is fully suppressed. This is not to say that the As--As bond plays no role; the inset of Fig.~\ref{fig:dft_calculations}(c) shows that, if one fixes the FeAs plane geometry of the collapsed 0.958$V_{0}^{\text{exp}}$ structure and artificially increases the interlayer As--As distance, then this will restore the magnetic phase \footnote{This only holds for volumes near the CT phase transition. At smaller volumes, such as 0.904$V_{0}^{\text{exp}}$, increasing the As--As interlayer distance does not restore magnetism.}. So, what we've found is that the As--As bond works against magnetism and lowers the critical pressure compared to an isolated FeAs plane, but its formation isn't necessary or sufficient to \textit{drive} the transition to the CT state.

So what is the nature of the CT phase?
We've established that the phase transition occurs when magnetism is fully suppressed, causing the FeAs planes to constrict, and that there is a direct connection between the stability of magnetism and the Fe--As bond geometry. As discussed earlier, the mechanism determining the equilibrium Fe--As bond geometry was identified in Ref.~\onlinecite{Belashchenko_2008} as a competition between covalent bonding (disfavoring magnetism) and exchange splitting (favoring magnetism) of the hybridized Fe $3d$ and As $4p$ states. Hence, the CT phase should be viewed as a fully nonmagnetic, \textit{strong covalent phase} that manifests due to covalency winning out in the Fe--As bonds \cite{diehl_2014} with increasing pressure.


Understanding that the CT phase is the product of a strong covalent Fe--As bond that fully suppresses magnetism offers insight on other results in the literature. First, a 122 pnictide is not in the CT phase if magnetism coexists with an As--As interlayer distance that is below 3 {\AA} (an example is applying pressure to $33\%$ Sr-doped {\CFA} \cite{jeffries_2014}). Second, there does not seem to be a requirement that the CT phase transition is first order. In the case of Sr-doped {\CFA}, according to our hydrostatic pressure measurements (see the Supplemental Materials \cite{SM_2016}) and Ref.~\onlinecite{knoner_2016}, the phase transition remains first order only when Sr doping remains low ($< 17.7\%$), while at larger dopings there is a sudden, yet continuous, increase in the As--Fe--As bond angle with increasing pressure as magnetism becomes suppressed. This kind of second-order phase transition behavior is also seen in BaFe$_{2}$As$_{2}$ \cite{Ba122CT_2011}, and so we conclude that 1) the CT phase is a general feature of the 122 family of pnictides, and 2) the critical pressure for the 300 K measurements in Ref.~\onlinecite{Ba122CT_2011} is determined by where the As--Fe--As bond angle plateaus, which is at 36 GPa instead of the quoted estimate of 27 GPa. Finally, it is worth noting that in rare-earth doped {\CFA} a superconducting state above 45 K emerges at the same time as a CT phase transition \cite{Saha_2012, CaPrPNAS_2011, CaPrSTM_2014}. Our results show that first order CT phase transitions are the result of a sudden quench of magnetism, which suggests that the CT phase in combination with a higher superconducting temperature are likely correlated in these materials. Further investigations in this direction are needed.

In summary, our FeK$\beta$ X-ray emission spectroscopy and DFT calculations establish the coexistence of local Fe moments with an interlayer As--As covalent bond with a length smaller than 3 {\AA} in Rh-doped Ca$_{0.8}$Sr$_{0.2}$Fe$_{2}$As$_{2}$. We find that the collapsed tetragonal phase is properly identified by a sudden constriction within the FeAs planes that occurs when magnetism is suppressed, which is due to covalent bonding between the hybridized Fe $3d$ and As $4p$ states winning out over exchange splitting. Therefore the collapsed tetragonal phase is not driven by forming an As--As bond and is instead a nonmagnetic and strongly covalent phase that should be distinguished from other magnetic or paramagnetic phases, even if under certain conditions they have relatively similar lattice parameters.

\acknowledgments
The authors would like to thank I.\ I.\ Mazin for helpful discussion and critical comments. K.\ Zhao thank Qinghua Zhang for the EDX measurement and the Alexander von Humboldt foundation for support. Financial support by the project DFG SPP1458 is acknowledged. J.\ Glasbrenner acknowledges the support of the NRC program at NRL. Parts of this research were carried out at the light source PETRA III (beamline P02.1) at DESY, a member of the Helmholtz Association (HGF). J.\ G.\ Cheng is supported by the MOST, NSFC, and CAS (Grant Nos. 2014CB921500, 11574377, XDB07020100, and QYZDB-SSW-SLH013). K.\ Zhao and J.\ K.\ Glasbrenner contributed equally to this work.

\end{document}